\documentclass[times, astrosymb]{aastex701}  
\pdfoutput=1 

\usepackage{amsmath,amstext}
\usepackage{graphicx}
\usepackage{booktabs}
\usepackage{hyperref}
\usepackage{enumitem}
\usepackage{subfig}

\shorttitle{Infrared Spectroscopy of L1415-IRS}
\shortauthors{Hillenbrand et al.}

\submitjournal{RNAAS}

\begin{document}

\title{The Post-Outburst Spectrum of L1415-IRS}
\correspondingauthor{Lynne A. Hillenbrand}

\author{Lynne A. Hillenbrand}
\affiliation{Department of Astronomy, California Institute of Technology, Pasadena, CA 91125, USA}
\email{lah@astro.caltech.edu}

\author[0000-0002-8963-8056]{Thomas P. Greene}
\affiliation{IPAC, Division of Physics, Mathematics, and Astronomy, California Institute of Technology, Pasadena, CA 91125, USA}
\email{tpgreene@ipac.caltech.edu}

\author{Evan R. Portnoi}
\affiliation{Department of Astronomy, California Institute of Technology, Pasadena, CA 91125, USA}
\email{evanrossportnoi@gmail.com}

\author[0000-0001-6091-163X]{Bringfried Stecklum}
\affiliation{Th\"uringer Landessternwarte Tautenburg, Sternwarte 5, Tautenburg, 07778, Germany}
\email{stecklum@tls-tautenburg.de}

\begin{abstract}
We present near-infrared spectroscopic observations of L1415-IRS, 
a Class I type young stellar object 
that underwent a factor of $>$30  photometric outburst in the early 2000s. 
Ours is the first reported spectroscopy of this deeply embedded source. 
Both components of L1415-IRS present continuum plus $H_2$ emission. 
The southern component has additional weak \ion{H}{1} Br$\gamma$ and $CO$ emission. 
Based upon our spectrum and evidence of a declining lightcurve, 
we hypothesize that either: the outburst was not large enough and therefore hot enough 
to produce the FU Ori type absorption spectrum commonly interpreted as an accretion disk atmosphere,
or the source has already recovered from FU Ori-type equatorial accretion to re-establish T Tauri-type magnetospheric accretion.
\end{abstract}


\section{Introduction}\label{sec:intro}

Young stellar objects (YSOs) undergo variable accretion
over a range of amplitudes and timescales \citep{fischer2023}, with the more
extreme large-amplitude and long-lived accretion outburst events 
potentially important for stellar mass assembly. 
Discovery of newly outbursting YSOs, as well as long-term monitoring of mature outbursts,
has accelerated over the past decade due to dedicated 
photometric surveys conducted at optical, near-infrared, and mid-infrared wavelengths.

LDN 1415 \citep{lynds1962} in Camelopardalis 
harbors the mid-infrared source IRAS 04376+5413, also known as L1415-IRS. 
With spectral index $\alpha=0.9$, the SED is an extreme Class I type disk and overlaps with protostars \citep{grossschedl2019}. 
In the near-infrared, the object is resolved into a north-south pair 
of sources: 2MASS J04413594+5419168 and 2MASS J04413587+5419116, separated by about 5\arcsec.5.

\citet{stecklum2006} and \cite{stecklum2007}  announced that an arc-shaped optical nebula 
had appeared just north of the northern source,
and argued based on source morphology that the two near-infrared objects are not point sources 
but rather the scattered light lobes of a bipolar outflow, seen nearly edge-on. 
Also reported were two Herbig–Haro objects (30" north and 60" south) 
aligned with the outflow axis indicated by the nebular structure and the two near-infrared lobes.  
Comparison to prior survey images suggested that a brightening of 3.8 mag had occurred over five years, between 2001 and 2006.
Optical spectroscopy of the new spatially offset nebula showed a P Cygni profile in the H$\alpha$ line.
\cite{stecklum2007} estimated the pre-outburst luminosity at $0.16 (d/170 {\rm pc})^2\ L_\odot$
with the post-outburst luminosity thus about 33 times that.

\cite{pawade2010} reported VRI and JHK photometry between 2006-2009, 
and low-resolution optical spectroscopy. 
They discussed morphological evolution in the nebula as the central source evolved following outburst, 
and traced the photometric decline ($\sim$0.03 mag/month).
Compared to 1998 2MASS photometry, the source was bluer post-outburst, 
but subsequently drifted redward during the photometric decline.

\cite{singh2023} presented optical lightcurves between 2006-2022, 
and also analyzed the photometric decline.  
Their optical spectroscopy of the nebula exhibited continuum plus weak H$\alpha$ emission 
but lacked the P Cygni profile of the \cite{stecklum2007} spectrum, which they interpreted
as disappearance of strong wind signatures. 

\section{New Photometric and Spectroscopic Observations}

The  WISE$+$NEOWISE mid-infrared lightcurve of L1415 IRS (see Figure)
exhibits a steady decline ($0.054 \pm 0.004$ mag/year)  over 2010-2022. 
There are signs of a flattening in 2023 and 2024, however.
The source is not visible in other archival
photometric data sets at near-infrared and optical wavelengths. 

We obtained K-band spectra at Keck Observatory with MOSFIRE \citep{mclean2012} on 2022-11-11 UT, 
separately observing the northern (N) and southern (S) sources 
using standard A-B-B-A dithering with 480 sec total integration time.
Each appeared pointlike in the 0.\arcsec7 seeing (see Figure) with the S source slightly fainter.
The configurable slit unit was set to a 1.0\arcsec wide longslit, producing $R\,\approx\,3000$ spectra
over $\lambda\,=\,1.95 - 2.4\,\mu$m.  Data processing was with PypeIt \citep{Prochaska2020}. 

The two spectra (see Figure) have similar shapes, 
likely dominated by extinction local to the star-forming core.  
Dereddening by $A_V=29$ mag is needed to match an FU Ori near-infrared spectral slope, 
with less extinction required for cooler spectral templates.  
The N and S sources are somewhat different in their emission properties, however.
The S source has strong Br$\gamma$ ($W_\lambda= -3.15\pm0.06$) and 
weak CO emission in the 2-0 ($W_\lambda= -2.34\pm0.07$) and 3-1 transitions.
The N source is mainly continuum with possible weak Br$\gamma$ emission ($W_\lambda= -0.28\pm0.09$).
Both spectra show $H_2$ 1-0 emission e.g. 2.12 $\mu$m, with the N source stronger.

\section{Discussion}

The hypothesis that the two near-infrared sources are lobes of reflected light emanating 
from the same embedded object viewed edge-on is not fully supported by the spectra,
which have slightly different appearances.  
The N source shows neither obvious CO nor strong Br$\gamma$ in emission; 
it also has no evidence for absorption features such as \ion{Na}{1}, \ion{Ca}{1} or \ion{Mg}{1}, 
exhibiting essentially just a veiled continuum spectrum.
For the S source, CO emission plus strong Br$\gamma$ emission indicate 
hot gas in the inner disk \citep{carr1989} and likely rapid accretion \citep[e.g.][]{najita2007}.

Our spectroscopy is unable to establish either source as the origin of the 
photometric rise in the 2001-2006 time frame \citep{stecklum2007}.
If the S source, the weak emission-line spectrum could indicate
an object in recovery from an FU Ori accretion outburst, with
disk accretion through an equatorial boundary layer now yielding to 
magnetospheric accretion at high latitudes but at a rate that is still quite high.
If the N source, the continuum spectrum could also indicate an object in recovery from an
FU Ori accretion outburst, where the accretion rate has dropped enough that 
an accretion disk atmosphere is no longer being produced.

The long and shallow mid-infrared photometric decline of L1415 IRS is not dissimilar 
to the fading patterns of several FU Ori type sources (e.g. BBW 76, FU Ori itself). 
However, there is an interesting photocenter shift over time, with the NEOWISE position
drifting southward by $\sim2\arcsec$ over $\sim10$ years.
Employing the method of \cite{stecklum2025}, it can be determined that it is the N source
that has declined while the S source has remained steady in NEOWISE.


If indeed an FU Ori outburst, 
this would be a relatively low-luminosity ($\sim5 L_\odot$) and short-lived ($\sim$ 20 year) example.
Alternately, the source could have never been an FU Ori, and instead 
experienced a V1647 Ori type ``intermediate" accretion outburst. 

\begin{figure*}[ht]
\includegraphics[width=0.49\linewidth]{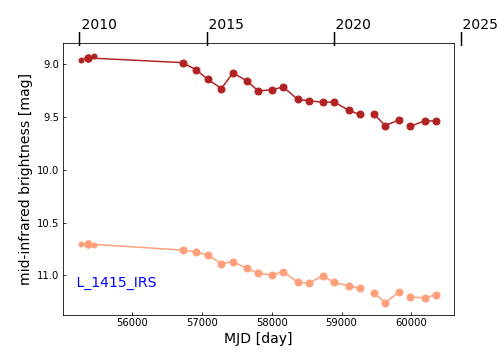}
\hspace{0.25truein}
\includegraphics[width=0.2\linewidth]{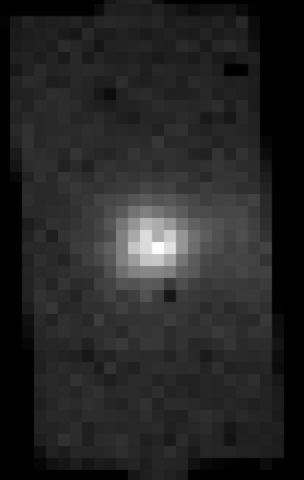}
\includegraphics[width=0.2\linewidth]{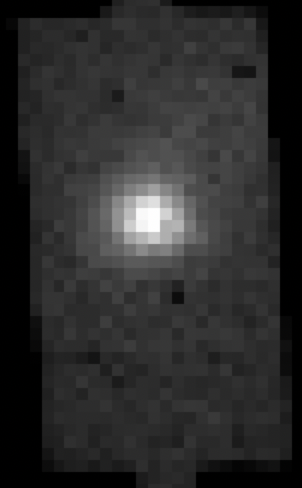}
\\
\includegraphics[width=0.49\linewidth]{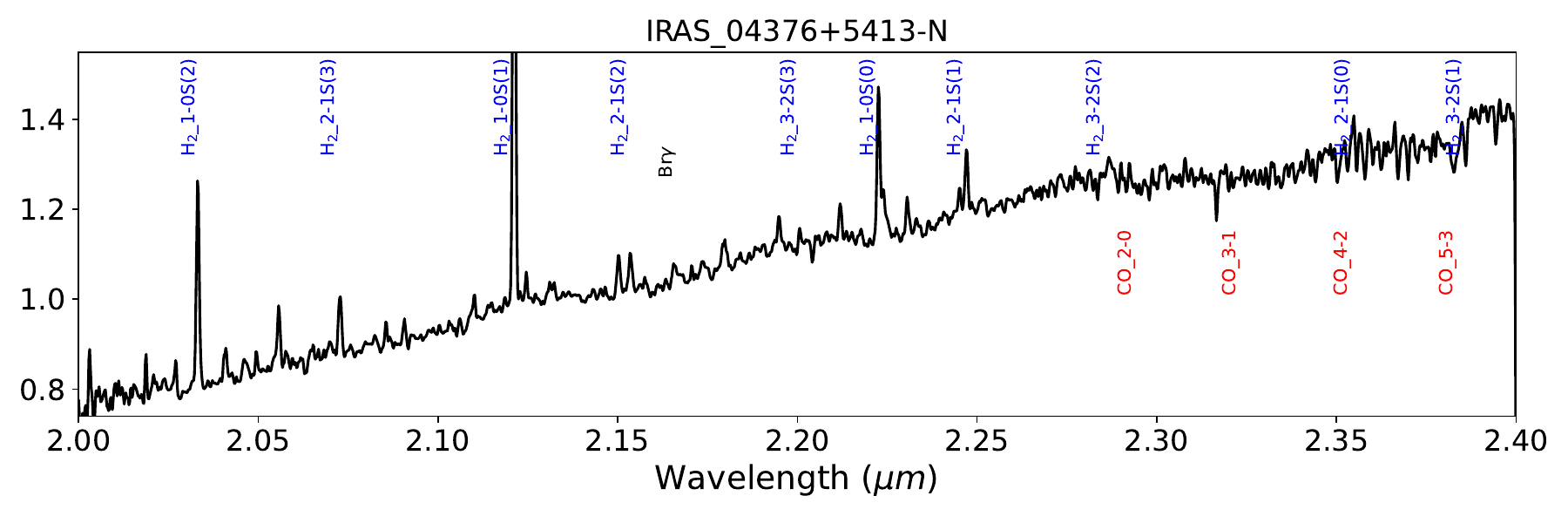}
\includegraphics[width=0.49\linewidth]{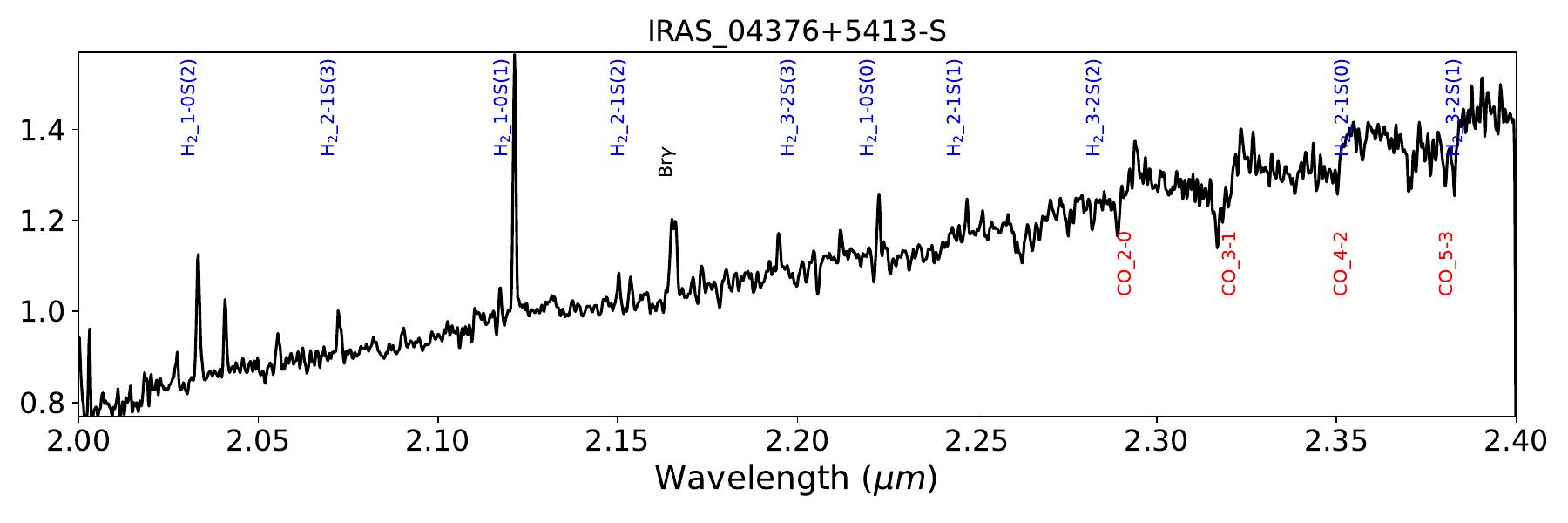}
\caption{
\textbf{Top left:} WISE$+$NEOWISE lightcurve of the unresolved source L1415-IRS.
\textbf{Top right:} Keck/MOSFIRE K-band images of the N (left) and S (right) pair.
\textbf{Bottom:} Keck/MOSFIRE spectra of IRAS 04376+5413 N (left) and S (right), 
gaussian-smoothed to reduce noise.
}
\label{fig}
\end{figure*}


\facility{Keck:I (MOSFIRE), NEOWISE}

\bibliography{bib.bib}{}
\bibliographystyle{aasjournal}

\end{document}